\documentclass[a4paper,nodate]{sab}  

\usepackage{graphicx} 
\usepackage{txfonts,natbib} 
\bibpunct{(}{)}{;}{a}{}{,} 

\usepackage[T1]{fontenc} 
\usepackage{t1enc} 
\usepackage{aeguill} 
\renewcommand\ackname{Acknowledgements.} 
\renewcommand\keywordname{Keywords.} 
\renewcommand\figureptname{Figure}
\renewcommand\tableptname{Table}

\renewcommand\noteaddname{Note added to proofs.}
\idline{Boletim da Sociedade Astron\^omica Brasileira, {\bf 29}, XX-XX}{1}

\begin{document} 
\title{Modelling the formation of the galactic bulge } 

\author{O. Cavichia \inst{1} and M. Moll\'a \inst{2}} 
\institute{Instituto de F\'isica e Qu\'imica, Universidade Federal de Itajub\'a, Av. BPS, 1303, 37500-903, Itajub\'a-MG, Brazil \email{cavichia@unifei.edu.br}  \and Departamento de Investigaci\'on B\'asica, CIEMAT, Avda. Complutense 40, E-28040 Madrid, Spain \email{mercedes.molla@ciemat.es}}

\Abstract {In this work we have assumed a Hernquist model with an inside-out formation for the Galactic bulge and, using a chemical evolution model, we were able obtain the bulge metallicity distribution function (MDF) for different radial regions. The preliminary results show that in the inner regions of the bulge the MDF has a higher fraction of metal poor stars, while this fraction is progressively diminished as moving outwards in the bulge. These results may explain the metallicity gradient observed in the Galactic bulge.}{Neste trabalho assumimos um modelo de Hernquist com uma forma\c{c}\~ao interna-externa para o bojo Gal\'actico e, usando um modelo de evolu\c{c}\~ao qu\'imica, fomos capazes de obter a distribui\c{c}\~ao de metalicidades do bojo (MDF) para diferentes regi\~oes radiais. Os resultados preliminares mostram que nas regi\~oes internas do bojo a MDF apresenta uma fra\c{c}\~ao maior de estrelas pobres em metais, enquanto que esta fra\c{c}\~ao diminui movendo-se para as regi\~oes externas do bojo. Estes resultados podem explicar o gradiente de metalicidades observado no bojo Gal\'actico.}


\maketitle 

\section{Introduction}
%
%
The Galactic bulge (GB) is the only galaxy bulge that can be resolved and can be studied with exquisite details. Hence, the bulge metallicity distribution function (MDF) can be traced for different regions within the bulge and can give us clues about the bulge formation scenario. Recently, observations (e.g. \citeauthor{zoccali08} 2008, \citeauthor{rojas14} 2014) have shown that the GB is composed of at least two components with different mean metallicities, and possibly different spatial distribution and kinematics. \citet{zoccali17} have demonstrated that the two components present a different spatial distribution, with the metal poor population being more centrally concentrated. There is no consensus in the literature regarding if this result point to a GB formation scenario formed by a bar instability or if the the GB formed from a spheroidal collapse. In this regarding, chemical evolution models (CEM) are important tools to understand the formation and evolution of the components of the Milky Way Galaxy (MWG) and other galaxies in the universe.  The outputs generated by the CEM are compared with the observational data and the physics adopted is changed self-consistently until satisfactory results are achieved. Precisely, one of the most important constraints for CEM is the metallicity distribution function (MDF). In particular, the MDF of a region is sensible to the time-scale that the region was formed. 

\begin{figure}[!ht]
\begin{center}
\includegraphics[width=8.cm]{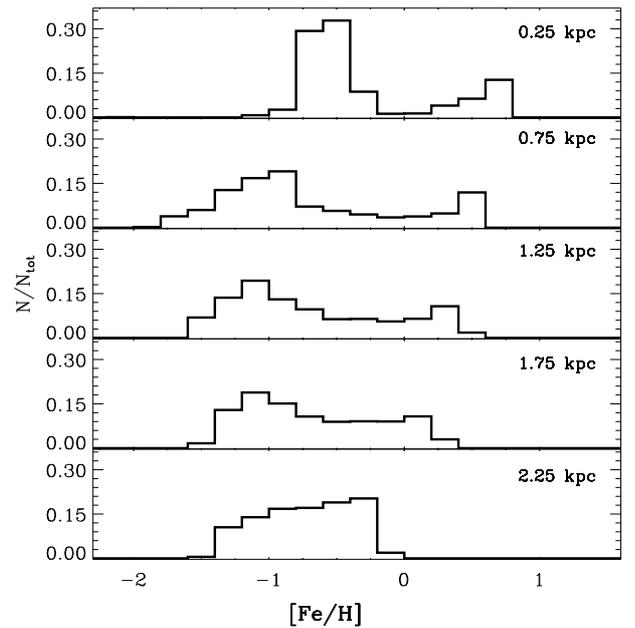}
\caption{Fe/H] histograms for different regions in the GB (0.25, 0.75, 1.25, 1.75 and 2.25 kpc), as labeled in each panel. The theoretical MDFs represent those from giant GB stars, being the stellar abundances computed taking into account the stellar life-times from Padova isochrones.\label{fig1}}
\end{center}
\end{figure}

\section{Model and results}

Our model is an update of the one from \citet{molla05} and the GB is formed assuming an inside-out growth scenario where the accretion of metal-poor gas from the halo builds the spheroidal component and also the disc. The infall rate for each radial region is given in \citet{molla16} and is obtained by imposing that after a Hubble time the system ends with a mass as observed for a Milky Way sized galaxy. In this work we have adopted the Hernquist mass model \citep{hernquist90} for the GB in order to constraint the scenario that formed the GB:
\begin{equation}
M(r)=M_{\rm b} \frac{r^2}{(r+a_{\rm b})^2},
\end{equation}
where $M_{\rm b}$ is the GB mass, adopted to be $1.85\times10^{10} {\rm M}_{\sun}$ (following \citeauthor{cavichia14} et al. 2014) and the characteristic scale-length $a_{\rm b}=r_{\rm eff} (1+\sqrt{2})^{-1}=0.124$ kpc using $r_{\rm eff}=300$~pc (bulge half-light radius) from \citet{wyse93}. We divided the GB in five concentric regions, being the innermost a sphere of 0.25 kpc radius. The subsequent regions are spherical shells with external radii 0.75, 1.25, 1.75 and 2.25 kpc, each one 0.5 kpc wide. The Hernquist mass distribution is obtained allowing gas infalling first in the innermost spherical region and, when the mass distribution in that region resembles that from the Hernquist profile at the same radius, the infall ceases at this spherical region and begins at the following spherical shell. The total collapse-time scale to form the GB is 2 Gyr. The resulting GB metallicity distribution function (MDF) for each spherical region is shown in Fig. \ref{fig1}, represented by the [Fe/H] histograms. Double peak histograms are observed for each region, with exception of the outermost one, suggesting a GB composed by two populations with different mean metallicities. Probably in the outermost spherical shell there is an overlap between the disc and bulge population and the two component population is not observed. Nonetheless, it is evident in this figure the differences in the MDFs for each region. In the innermost one the fraction of the metal poor population is higher than the metal rich one. However, as moving further out to the other spherical shells, the fraction of metal poor stars decreases and the one from metal rich population increases. In spite of that, the positions of the peaks do not change considerably, considering the shells 0.75, 1.25 and 1.75 kpc. This results are in agreement with the most recent observations from \citet{zoccali17}. 

\begin{figure}[!ht]
\begin{center}
\includegraphics[width=8.cm]{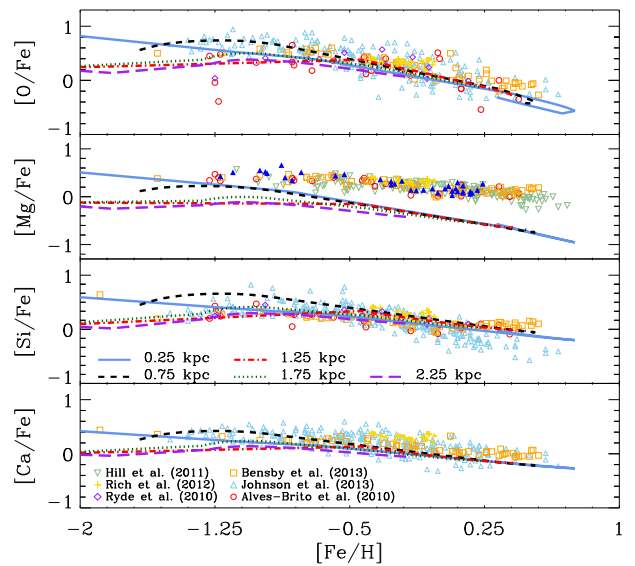}
\caption{Comparison between the predictions of our model at different radial regions and the literature data for $[\alpha/\mbox{Fe}]$  where $\alpha$ stands for O, Mg, Si, and Ca,  vs. [Fe/H]. The observational data for the bulge are labeled in the last panel (from top to bottom). The line codes of the models are as labeled in the third panel (from top to bottom). The observational data for the bulge are open circles from \citet{alves-brito10}, open squares from \citet{bensby13}, open upside triangles from \citet{johnson13}, open upside-down triangles from \citet{hill11}, open stars from \citet{rich12}, and open diamonds from \citet{ryde10}. \label{fig2}}
\end{center}
\end{figure}

In Fig.\ref{fig2} it is shown $[\alpha/{\rm Fe}]$ versus $[{\rm Fe}/{\rm H}]$, where $\alpha$ stands for the $\alpha$-elements: O, Mg, Si and Ca. Overall, our model is able to reproduce the observational data regarding the $[{\rm O}/{\rm Fe}]$, $[{\rm Si}/{\rm Fe}]$  and $[{\rm Ca}/{\rm Fe}]$ abundance ratios.  The $[{\rm Ca}/{\rm Fe}]$ abundance ratio of \citet{bensby13} and \citet{rich12} are higher than the results of the model. In our case, we adopted the yields of O and Mg for high mass stars from \citet{woosley95} and it is known that these yields needed to be increased in order to reproduce the most recent data of the bulge (\citeauthor {cavichia14} et al. 2014 and \citeauthor{molla15} et al. 2015).  Interesting, the model predicts different $[\alpha/{\rm Fe}]$ ratios for each spherical region, the ratio being higher at the innermost regions (0.25 and 0.75 kpc) and lower at the outermost regions, especially at low $[{\rm Fe}/{\rm H}]$ abundances. Higher $[\alpha/{\rm Fe}]$ ratios at low [Fe/H] indicate a fast enrichment by massive stars  which explode as type II  supernovae. On the other hand, at higher metallicities ([Fe/H]$ > -0.5$~dex) the differences in $[\alpha/{\rm Fe}]$ ratios of the models are less important, probably because at these metallicities the type Ia supernovae have enriched the interstellar medium.  

\section{Discussion}

The results of Fig. \ref{fig2} suggest that the spread in $[\alpha/{\rm Fe}]$ ratios, as shown by the observational data at lower metallicities, may be the result of mixing stars from different radial locations within the GB. Since the differences are in the range of $\sim$ 0.30 dex, it will be very difficult to detect this trend given the level of the uncertainties of the present observations. To confirm this possible gradient of $[\alpha/{\rm Fe}]$ in the GB predicted by our models, more data with higher precision is needed, mainly for lower metallicities and galactocentric distances.

\begin{acknowledgements} 
This work has made use of the computing facilities available at the Laboratory of Computational Astrophysics of the Universidade Federal de Itajub\'a (LAC-UNIFEI). The LAC-UNIFEI is maintained with grants from CAPES, CNPq and FAPEMIG. O.C. would like to thank CAPES, PGF/UNIFEI and CIEMAT.
\end{acknowledgements}

\end{document}